\documentclass[12pt, a4paper, onecolumn]{article}

\usepackage{fullpage}

\usepackage[english]{babel}
\usepackage{amsmath}         
\usepackage{amssymb}         

\usepackage{graphicx}
\usepackage{url}           

\long\def\symbolfootnote[#1]#2{
\begingroup
	\def\thefootnote{\fnsymbol{footnote}}\footnote[#1]{#2}
\endgroup}

\begin{document}

\title{Using Rank Aggregation for Expert Search in Academic Digital Libraries}

\author{Catarina Moreira\\ \small \texttt{catarina.p.moreira@ist.utl.pt}\\
\and
Bruno Martins\\ \small \texttt{bruno.g.martins@ist.utl.pt}
\and
P\'{a}vel Calado\\ \small \texttt{pavel.calado@ist.utl.pt}
\and
\\Instituto Superior T\'{e}cnico, INESC-ID\\ Av. Professor Cavaco Silva, 2744-016 Porto Salvo, Portugal\\  
\\ \small Published in Proceedings of the 3rd Simp\'{o}sio de Inform\'{a}tica, 2011, Portugal\\
}

\date{}
\maketitle

\begin{abstract}
The task of expert finding has been getting increasing attention in information retrieval literature. However, the current state-of-the-art is still lacking in principled approaches for combining different sources of evidence. This paper explores the usage of unsupervised rank aggregation methods as a principled approach for combining multiple estimators of expertise, derived from the textual contents, from the graph-structure of the citation patterns for the community of experts, and from profile information about the experts. We specifically experimented two unsupervised rank aggregation approaches well known in the information retrieval literature, namely CombSUM and CombMNZ. Experiments made over a dataset of academic publications for the area of Computer Science attest for the adequacy of these methods.
\end{abstract}

\symbolfootnote[0]{This work was partially supported by the ICP Competitiveness and Innovation Framework Program of the European Commission, through the European Digital Mathematics Library (EuDML) project -- \url{http://www.eudml.eu/}}

\section{Introduction}

The automatic search for knowledgeable people in the scope of specific user communities, with basis on documents describing people's activities, is an information retrieval problem that has been receiving increasing attention~\cite{Pavel09Search}. Usually referred to as \emph{expert finding}, the task involves taking a short user query as input, denoting a topic of expertise, and returning a list of people sorted by their level of expertise in what concerns the query topic. 

Several effective approaches for finding experts have been proposed, exploring different retrieval models and different sources of evidence for estimating expertise. However, the current state-of-the-art is still lacking in principled approaches for combining the multiple sources of evidence that can be used to estimate expertise. 

More recently, several authors have also proposed unsupervised learning to rank methods, based on rank aggregation approaches originally proposed in areas such as statistics or the social sciences~\cite{Pine11Data,Klementiev09Unsupervised}. This paper explores the usage of unsupervised rank aggregation methods in the expert finding task, specifically combining a large pool of estimators for expertise. These include estimators derived from the textual similarity between documents and queries, from the graph-structure of the citation patterns for the community of experts, and from profile information about the experts. We have built a prototype expert finding system using rank aggregation methods, and evaluated it on an academic publications dataset from the Computer Science domain.

The rest of this paper is organized as follows: Section 2 presents the main concepts and related works. Section 3 presents the rank aggregation approaches used in our experiments. Section 4 introduces the multiple features upon which we leverage for estimating expertise. Section 5 presents the experimental evaluation of the proposed methods, detailing the datasets and the evaluation metrics, as well as the obtained results. Finally, Section 6 presents our conclusions and points directions for future work.

\section{Concepts and Related Work}

Serdyukov and Macdonald have surveyed the most important concepts and representative previous works in the expert finding task~\cite{Pavel09Search,Macdonald08Voting}. Two of the most popular and well-performing types of methods are the profile-centric and the document-centric approaches~\cite{Craswell06Overview,Soboroff07Overview}. Profile-centric approaches build an expert profile as a pseudo document, by aggregating text segments relevant to the expert~\cite{Balog06Formal}. These profiles are latter indexed and used to support the search for experts on a topic. 
Document-centric approaches are typically based on traditional document retrieval techniques, using the documents directly. In a probabilistic approach to the problem, the first step is to estimate the conditional probability $p(q|d)$ of the query topic $q$ given a document $d$. Assuming that the terms co-occurring with an expert can be used to describe him, $p(q|d)$ can be used to weight the co-occurrence evidence of experts with $q$ in documents. The conditional probability $p(c|q)$ of an expert candidate $c$ given a query $q$ can then be estimated by aggregating all the evidences in all the documents where $c$ and $q$ co-occur. Experimental results show that document-centric approaches usually outperform profile-centric approaches~\cite{Soboroff07Overview}. 

Many different authors have proposed sophisticated probabilistic retrieval models, specific to the expert finding task, with basis on the document-centric approach~\cite{Balog06Formal,Petkova07Proximity,Pavel09Search}. For instance Cao et al. proposed a two-stage language model combining document relevance and co-occurrence between experts and query terms~\cite{Cao06Research}. Fang and Zhai derived a generative probabilistic model from the probabilistic ranking principle and extend it with query expansion and non-uniform candidate priors~\cite{Fang07Probabilistic}. Zhu et al. proposed a multiple window based approach for integrating multiple levels of associations between experts and query topics in expert finding~\cite{Zhu07Open}. More recently, Zhu et al. proposed a unified language model integrating many document features for expert finding~\cite{Zhu08Modeling}. Although the above models are capable of employing different types of associations among query terms, documents and experts, they mostly ignore other important sources of evidence, such as the importance of individual documents, or the co-citation patterns between experts available from citation graphs. In this paper, we offer a principled approach for combining a much larger set of expertise estimates.

In the Scientometrics community, the evaluation of the scientific output of a scientist has also attracted significant interest due to the importance of obtaining unbiased and fair criteria. Most of the existing methods are based on metrics such as the total number of authored papers or the total number of citations. A comprehensive description of many of these metrics can be found in~\cite{Sidiropoulos05citation,Sidiropoulos06generalized}. Simple and elegant indexes, such as the Hirsch index, calculate how broad the research work of a scientist is, accounting for both productivity and impact. Graph centrality metrics inspired on PageRank, calculated over citation or co-authorship graphs, have also been extensively used~\cite{Liu05authorship}. In ~\cite{Moreira11MSc}, the reader can find a full survey of different expert finding approaches in the literature.

Previous studies have addressed the problem of combining multiple information retrieval mechanisms through unsupervised rank aggregation, often with basis on methods that take their inspiration on voting protocols proposed in the area of statistics and in the social sciences. Given $M$ voters (i.e., the different estimators of expertise) and $N$ objects (i.e., the experts), we can see each voter as returning an ordered list of the $N$ objects according to their own preferences. From these $M$ ordered lists, the problem of unsupervised rank aggregation concerns with finding a single consensus list which optimally combines the $M$ rankings. There are different methods for addressing the problem which, according to Julien Ah-Pine~\cite{Pine11Data}, can be divided into two large families of methods:

\begin{itemize}
\item {\bf Positional methods} - For each object, we consider the preferences (i.e., the scores) given by each voter, aggregating them through some particular technique and finally re-ranking objects using the aggregated preferences. The first positional method was proposed by Borda, but linear and non-linear combinations of preferences, such as arithmetic mean or the triangular norm, are also frequently used~\cite{Fox94Combination,Pine11Data}. 

\item {\bf Majoritarian methods} - Pairwise comparison matrices are computed for the objects, mostly based upon the aggregation of order relations using association criteria such as Condorcet's criterion, or distance criteria such as Kendall's distance. Other majoritarian methods have also recently been proposed, using Markov chain models~\cite{Dwork01Rank} or techniques from multicriteria decision theory~\cite{Farah07Outranking}.
\end{itemize}

Fox and Shaw~\cite{Fox94Combination,Pine11Data} defined several rank aggregation techniques (e.g., CombSUM and CombMNZ) which have been the object of much IR research since, including in the area of expert search~\cite{Macdonald08Voting}. In our experiments, we compared the CombSUM and CombMNZ unsupervised rank aggregation methods, which are detailed in Section 3.

\section{Rank Aggregation for Expert Retrieval}

Given a set of queries $Q = \{q_1,\ldots, q_{|Q|}\}$ and a collection of candidate experts $E = \{e_1,\ldots, e_{|E|}\}$, each associated with specific documents describing his topics of expertise, a testing corpus consists of a set of query-expert pairs, each $(q_i, e_j) \in Q \times E$, upon which a relevance judgment indicating the match between $q_i$ and $e_j$ is assigned by a labeler. This relevance judgment can be a binary label, e.g., relevant or non-relevant, or an ordinal rating indicating relevance, e.g., definitely relevant, possibly relevant, or non-relevant. For each instance $(q_i, e_j)$, a feature extractor produces a vector of features that describes the match between $q_i$ and $e_j$. Features can range from classical IR estimators computed from the documents associated with the experts (e.g., term frequency, inverse document frequency, BM25, etc.) to link-based features computed from networks encoding relations between the experts in $E$ (e.g., PageRank). The inputs of an unsupervised rank aggregation algorithm comprise a set of query-expert pairs corpus, their corresponding feature vectors, and the corresponding relevance judgments. The output produces a ranking score resulting from the aggregation of the multiple features. The relevance of each expert $e_j$ towards the query $q$ is determined through this aggregated score. In this paper, we experimented with the CombSUM and CombMNZ approaches.

The CombSUM and CombMNZ unsupervised rank aggregation algorithms were originally proposed by Fox and Shaw~\cite{Fox94Combination}. These algorithms are used to aggregate the information gathered from different sources (i.e., different features) in order to achieve more accurate ranking results than using individual scores. Both CombSUM and CombMNZ use normalized sums for the different features. To perform the normalization, we applied the Min-Max Normalization procedure, which is given by Equation~\ref{eq:norm}.
\begin{equation}
NormalizedValue = \frac{Value - minValue}{maxValue - minValue}
\label{eq:norm}
\end{equation}

The CombSUM score of an expert $e$ for a given query $Q$ is the sum of the normalized scores received by the expert in each individual ranking, and is given by Equation~\ref{eq:CombSUM}.
\begin{equation}
CombSUM(e,Q) = \sum_{j = 1} ^k score_{j}(e, Q)
\label{eq:CombSUM}
\end{equation}

Similarly, the CombMNZ score of an expert $e$ for a given query $Q$ is defined by Equation~\ref{eq:CombMNZ}, where $r_e$ is the number of non-zero similarities.
\begin{equation}
CombMNZ(e,Q) = CombSUM(e, Q)\times r_e
\label{eq:CombMNZ}
\end{equation}

\section{Features for Estimating Expertize}\label{features}

The considered set of features for estimating the expertize of a given researcher towards a given query can be divided into three groups, namely textual features, profile features and network features. The textual features are similar to those used in standard text retrieval systems (e.g., TF-IDF and BM25 scores). The profile similarity features correspond to importance estimates for the authors, derived from their profile information (e.g., number of papers published). Finally, the network features correspond to importance and relevance estimates computed from the author co-authorship and co-citation graphs.

\subsection{Textual Similarity Features}

To build some of our estimators of expertise, we used the textual similarity between the query and the contents of the documents associated to the candidate experts. In the domain of academic digital libraries, the associations between documents and experts can easily be obtained from the authorship information. For each topic-expert pair, we used the $Okapi BM25$ document-scoring function, to compute the textual similarity features. Okapi BM25 is a state-of-the-art IR ranking mechanism composed of several simpler scoring functions with different parameters and components (e.g., term frequency and inverse document frequency). It can be computed through the formula in Equation~\ref{eq:bm25}, where ${\it Terms}(q)$ represents the set of terms from query {\it q}, ${\it Freq(i,d)}$ is the number of occurrences of term {\it i} in document $d$, $|d|$ is the number of terms in document $d$, and $\mathcal{A}$ is the average length of the documents in the collection. The values given to the parameters $k_1$ and $b$ were 1.2 and 0.75 respectively. Most previous IR experiments use these default values for the $k_1$ and $b$ parameters.

\begin{equation} \label{eq:bm25}
\begin{split}
{BM25(q,d)} = \sum_{i \in Terms(q)}\log \left( \frac{N-Freq(i)+0.5}{Freq(i)+0.5} \right) \times\\ \frac{(k_1+1) \times \frac{Freq({i,d})}{|d|}} {\frac{Freq({i,d})}{|d|} + k_1 \times (1 - b + b \times \frac{|d|}{\mathcal{A}})}
\end{split}
\end{equation} 
We also experimented with other textual features commonly used in ad-hoc IR systems, such as {\it Term Frequency} (TF) and {\it Inverse Document Frequency} (IDF).

Term Frequency (TF) corresponds to the number of times that each individual term in the query occurs in all the documents associated with the author. Equation~\ref{eq:tf} describes the TF formula, where $i \in Terms(q)$ represents the set of terms from query {\it q}, $j \in Docs(a)$ is the set of documents having {\it a} as author, $Freq(i, d_{j})$ is the number of occurrences of term {\it i} in document $d_{j}$ and $\left|d_{j}\right|$ represents the number of terms in document $d_{j}$.

	\begin{equation} 
TF_{q,a} =  \sum_{j \in Docs(a)} \sum_{i \in Terms(q)} \frac{Freq({i,d_j})}{|d_j|}
\label{eq:tf}
\end{equation}

The Inverse Document Frequency (IDF) corresponds to the sum of the values for the inverse document frequency of each query term and is given by Equation~\ref{eq:idf}. In this formula, $|D|$ is the size of the document collection and $f_{i, D}$ corresponds to the number of documents in the collection where the $i_{th}$ query term occurs.
		
\begin{equation} 
	\label{eq:idf}
	IDF_q = \sum_{i \in Terms(q)} \log \frac{|D|}{f_{i, D}}
\end{equation}

We also used other simpler features such as the number of unique authors associated with documents containing the query topics, the range of years since the first and last publications of the author containing the query terms and the document length.

In the computation of the textual features, we considered two different sources of evidence extracted from the documents, namely (i) a stream consisting of the titles, and (ii) a stream using the abstracts of the articles. Separate features were computed for each of these streams.

\subsection{Profile Information Features}

We also considered a set of profile features related to the amount of published materials associated with authors, generally taking the assumption that highly prolific authors are more likely to be considered experts. Most of the features based on profile information are query independent, meaning that they have the same value for different queries. The considered set of profile features are based on the number of publications in conferences and in journals with and without the query topics in their contents, the average number of papers and articles per year, and the temporal interval between the first and the last publications.

\subsection{Co-citation and Co-authorship Features}

Scientific impact metrics computed over scholarly networks, encoding co-citation and co-authorship information, can offer effective approaches for estimating the importance of the contributions of particular publications. Thus, we considered a set of features that estimate expertise with basis on co-citation and co-authorship information. The considered features are divided in two sets, namely (i) citation counts and (ii) academic indexes. Regarding citation counts, we used the total, the average and the maximum number of citations of papers containing the query topics, the average number of citations per year of the papers associated with an author and the total number of unique collaborators which worked with an author. 

Regarding academic impact indexes, we used the following features:

\begin{itemize}
\item {\bf Hirsch index} of the author and of the author's institution, measuring both the scientific productivity and the scientific impact of the author or the institution~\cite{Hirsch05Index}. A given author or institution has an Hirsch index of $h$ if $h$ of his $N_p$ papers have at least $h$ citations each, and the other $(N_p - h)$ papers have at most $h$ citations each. Authors with a high Hirsch index, or authors associated with institutions with a high Hirsch index, are more likely to be considered experts.
\item The {\bf $h$-$b$-index}, which extends the Hirsch index for evaluating the impact of scientific topics in general~\cite{Banks06extension}. In our case, the scientific topic is given by the query terms and thus the query has an $h$-$b$-index of $i$ if $i$ of the $N_p$ papers containing the query terms in the title or abstract have at least $i$ citations each, and the other $(N_p - i)$ papers have at most $i$ citations each.

\item {\bf Contemporary Hirsch index} of the author, which adds an age-related weighting to each cited article, giving less weight to older articles~\cite{Antonis06Generalized}. A researcher has a contemporary Hirsch index $h^c$ if $h^c$ of his $N_p$ articles have a score of $S^c(i) >= h^c$ each, and the rest $(N_p - h^c)$ articles have a score of $S^c(i) <= h^c$. For an article $i$, the score $S^c(i)$ is defined as:
\begin{equation}
S^c(i) = \gamma * (Y(now) - Y(i) + 1)^{-\delta} * |CitationsTo(i)|
\end{equation}
In the formula, $Y(i)$ refers to the year of publication for article $i$. The $\gamma$ and $\delta$ parameters are set to $4$ and $1$, respectively, meaning that the citations for an article published during the current year account four times, the citations for an article published 4 years ago account only one time, the citations for an article published 6 years ago account $4/6$ times, and so on.

\item {\bf Trend Hirsch index}~\cite{Antonis06Generalized} for the author, which assigns to each citation an exponentially decaying weight according to the age of the citation, this way estimating the impact of a researcher's work in a particular time instance. A researcher has a trend Hirsch index $h^t$ if $h^t$ of his $N_p$ articles get a score of $S^t(i) >= h^t$ each, and the rest $(N_p - h^t)$ articles get a score of $S^t(i) <= h^t$. For an article $i$, the score $S^t(i)$ is defined as shown bellow:
\begin{equation}
S^t(i) = \gamma * \sum_{\forall x \in C(i)} (Y(now) - Y(x) + 1)^{-\delta}
\end{equation}
Similarly to the case of the contemporary Hirsch index, the $\gamma$ and $\delta$ parameters are here also set to $4$ and $1$, respectively.

\item {\bf Individual Hirsch index} of the author, computed by dividing the value of the standard Hirsch index by the average number of authors in the articles that contribute to the Hirsch index of the author, in order to reduce the effects of frequent co-authorship with influential authors~\cite{Batista06Possible}.

\item The {\bf $a$-index} of the author or the author's institution, measuring the magnitude of the most influential articles. For an author or an institution with an Hirsch index of $h$ that has a total of $N_{c,tot}$ citations toward his papers, we say that he has an $a$-index of $a = N_{c,tot} / h^2$.

\item The {\bf $g$-index} of the author or his institution, also quantifying scientific productivity with basis on the publication record~\cite{Egghe06Theory}. Given a set of articles associated with an author or an institution, ranked in decreasing order of the number of citations that they received, the g-index is the (unique) largest number such that the top $g$ articles received on average at least $g$ citations.

\item The {\bf $e$-index} of the author~\cite{ZhangEIndex} which represents the excess amount of citations of an author. The motivation behind this index is that we can complement the $h$-index by taking into account these excess amounts of citations which are ignored by the $h$-index.
The $e$-index is given by the Equation~\ref{eq:e-index}:
\begin{equation}
e = \sum_{j=1}^{h} \sqrt{cit_j-h^2}
\label{eq:e-index}
\end{equation}
In the above equation, $cit_j$ are the citations received by the $j_th$ paper and $h$ is the $h$-index.

\end{itemize}

We also followed the ideas of Chen et al.~\cite{Chen07Finding} by considering a set of network features that estimate the influence of individual authors using PageRank, a well-known graph linkage analysis algorithm that was introduced by the Google search engine~\cite{Brin99PageRank}. PageRank assigns a numerical weighting to each element of a linked set of objects (e.g., hyperlinked Web documents or articles in a citation network) with the purpose of measuring its relative importance within the set. The PageRank value of a node is defined recursively and depends on the number and PageRank scores of all other nodes that link to it (i.e., the incoming links). A node that is linked to by many nodes with high PageRank receives a high rank itself. 

Formally, given a graph with $N$ nodes $i=1,2,\cdots,N$, with $L$ directed links that represent references from an initial node to a target node with weights $\alpha=1,2,\cdots,L$, the PageRank $Pr_i$ for the $i$th node is defined by:

\begin{equation}
Pr_i = \frac{0.5}{N} + 0.5 \sum_{j \in inlinks(L,i)} \frac{\alpha_j Pr_j}{outlinks(L,j)}
\end{equation}

In the formula, the sum is over the neighboring nodes $j$ in which a link points to node $i$. The first term represents the random jump in the graph, giving a uniform injection of probability into all nodes in the graph. The second term describes the propagation of probability corresponding to a random walk, in which a value at node $j$ propagates to node $i$ with probability $\frac{\alpha_j Pr_j}{outlinks(L,j)}$. 

The PageRank-based features that we considered correspond to the sum and average of the PageRank values associated to the papers of the author that contain the query terms, computed over a directed graph representing citations between papers. Each citation link in the graph is given a score of $1/N$, where $N$ represents the number of authors in the paper. Authors with high PageRank scores are more likely to be considered experts.

\section{Experimental Validation}

The main hypothesis behind this work is that unsupervised rank aggregation approaches can be effectively used in the context of expert search tasks, in order to combine different estimators of relevance in a principled way, this way improving over the current state-of-the art. To validate this hypothesis, we have built a prototype expert search system, using two unsupervised rank aggregation methods, namely the CombSUM and CombMNZ methods.

We implemented the methods responsible for computing the features listed in the previous section, using the {\it Microsoft SQL Server 2008} relational database (e.g., the full-text search capabilities for computing the textual similarity features) together with existing Java software packages (e.g., the LAW\footnote{\url{http://law.dsi.unimi.it/software.php}} package for computing PageRank). 

The validation of the prototype required a sufficiently large repository of textual contents describing the expertise of individuals within a specific area. In this work, we used a dataset for evaluating expert search in the Computer Science research domain, corresponding to an enriched version of the DBLP\footnote{\url{http://www.arnetminer.org/citation}} database made available through the Arnetminer project. DBLP data has been used in several previous experiments regarding citation analysis~\cite{Sidiropoulos05citation,Sidiropoulos06generalized} and expert search~\cite{Deng08Formal}. It is a large dataset covering both journal and conference publications, and where substantial effort has been put into resolving the problem of author identity resolution, i.e., references to the same persons with other names. 

Table~\ref{t1} provides a statistical characterization for the DBLP dataset. In this dataset, we have a large collection of articles with a large number of citations between them, but more than half of the articles have no abstracts associated to them. Thus, it would be expected for textual similarity features to not perform particularly well. 

\begin{table}[th!]
\resizebox{\columnwidth}{!} {
\begin{tabular}{ l c}
  
Dataset
  Property																																		& Value								\\
\hline

 Total Authors 																															& ~~~~1 033 050~~~~		\\
 Total Publications 																													& ~~~~1 632 440~~~~ 	\\
 Total Publications containing Abstract																			& ~~~~653 514~~~~ 		\\
 Total Papers Published in Conferences~~~~~~~~~~~~~~~~~~~~~~~~~~~~~~~~~~~~~~ & ~~~~606 953~~~~			\\
 Total Papers Published in Journals 																					& ~~~~436 065~~~~ 		\\
 Total Number of Citations Links 																						& ~~~~2 327 450~~~~ 	\\
  \hline
\end{tabular}
}
\caption{Statistical characterization for the DBLP dataset used in our experiments.}
\label{t1}
\end{table}

To validate the different learning to rank methods, we also needed a set of queries with the corresponding author relevance judgments. We used the relevant judgments provided by Arnetminer\footnote{\url{http://arnetminer.org/lab-datasets/expertfinding/}} which have already been used in other expert finding experiments~\cite{yang09bole}. The Arnetminer dataset comprises a set of 13 query topics, each associated to a list of expert authors. 

In order to add negative relevance judgments (i.e., complement the dataset with unimportant authors for each of the query topics), we searched the dataset with the keywords associated to each topic, retrieving the top $n/2$ authors according to the BM25 metric and retrieving $n/2$ authors randomly selected from the dataset, where $n$ corresponds to the number of expert authors associated to each particular topic. Table~\ref{judgments} shows the distribution for the number of experts associated to each topic in the collection.
\begin{table}[th!]
\resizebox{\columnwidth}{!} {
\begin{tabular}{l c l c}

~~{\bf Query Topics}			& ~~{\bf Authors}		&~~{\bf Query Topics} 				&~~{\bf Authors}	\\
\hline

 ~~Boosting (B)									& ~~46									 				&	~~Natural Language (NL)					& ~~41						\\
 ~~Computer Vision (CV)					&	~~176									 				&	~~Neural Networks	 (NN)					& ~~103						\\
 ~~Cryptography (C)							&	~~148									 				& ~~Ontology				 (O)					& ~~47						\\
 ~~Data Mining (DM)							&	~~318									 				& ~~Planning				 (P)					& ~~23						\\
 ~~Information Extraction (IE)	&	~~20									 				& ~~Semantic Web		 (SW)					& ~~326						\\
 ~~Intelligent Agents	(IA)			&	~~30									 				& ~~Support Vector Machines (SVM)	& ~~85						\\
 ~~Machine Learning	(ML)				& ~~34									 				& 																&								\\
\hline
\end{tabular}
}
\caption{Characterization of the Arnetminer dataset of Computer Science experts.}
\label{judgments}
\end{table}

To measure the quality of the results produced by the different rank aggregation algorithms, we used two different performance metrics, namely the Precision at $k$ (P@k) and the Mean Average Precision (MAP).

Precision at rank $k$ is used when a user wishes only to look at the first $k$ retrieved domain experts. The precision is calculated at that rank position through Equation~\ref{eq:PrecisionRank}.
\begin{equation}
P@k=\frac{r\left(k\right)}{k}
\label{eq:PrecisionRank}
\end{equation}
In the formula, $r(k)$ is the number of relevant authors retrieved in the top {\it k} positions. $P@k$ only considers the top-ranking experts as relevant and computes the fraction of such experts in the top-$k$ elements of the ranked list.

The Mean of the Average Precision over test queries is defined as the mean over the precision scores for all retrieved relevant experts. It is given by:

\begin{equation}
MAP[r] := \frac{\sum_{k=1}^n P@k[r] \times I\{ g_{r_k} = \max(g) \}}{\sum_{k=1}^n I\{ g_{r_k} = \max(g) \}}  
\end{equation}

As before, $n$ is the number of experts associated with query $q$. In the case of our datasets, $\max(g) = 1$ (i.e., we have 2 different grades for relevance, 0 or 1).

Table~\ref{t2} presents the obtained results over the dataset, when considering the complete set of features described in Section~\ref{features}. The obtained results attest for the adequacy of both unsupervised rank aggregation approaches, showing that CombSUM and CombMNZ achieve a similar performance, with CombMNZ slightly outperforming CombSUM, in terms of MAP. 
\begin{table*}[th!]
\resizebox{\columnwidth}{!} {
\begin{tabular}{ l c c c c c c }

				 											&~~~~P@5~~~~ 		& ~~~~P@10~~~~ 	& ~~~~P@15~~~~ 	& ~~~~P@20~~~~ 	& ~~~~MAP~~~~ 	\\
\hline

~~CombSUM~~~~~~~~~~~~~				& 0.5076					& 0.4846				& 0.4769				& 0.5115 				& 0.5266  \\
~~CombMNZ 										& {\bf 0.6000}		& {\bf 0.6077}	& {\bf 0.6141}	& {\bf 0.6256}	& {\bf 0.5832} \\
\hline

\end{tabular}
}
\caption{Results of the CombSUM and CombMNZ methods.}
\label{t2} 
\end{table*}
In a separate experiment, we attempted to measure the impact of the different types of ranking features on the quality of the results. Using the best performing rank aggregation algorithm, namely the CombMNZ method, we separately measured the results obtained by using approaches that considered (i) only the textual similarity features, (ii) only the profile features, (iii) only the network features, (iv) textual similarity and profile features, (v) textual similarity and network features and (vi) profile and network features. Table~\ref{t3} shows the obtained results, where we also compare them with the previous results reported by Yang et al.~\cite{yang09bole} for their supervised approach for expert finding. 
\begin{table}[th!]
\resizebox{\columnwidth}{!} {
\begin{tabular}{ l c c c c c}

																			& ~~~~P@5~~~~		& ~~~~P@10~~~~	& ~~~~P@15~~~~	& ~~~~P@20~~~~	& ~~~~MAP~~~~	\\
\hline
Text Similarity + Profile + Network~~  & 0.6000					& 0.6077				& 0.6141				& 0.6256				& 0.5832 \\
Text Similarity + Profile							 & 0.5231		 			& 0.5615  			& 0.5487  			& 0.5577				& 0.5469 		\\
Text Similarity + Network					   	 & 0.5538 				& 0.5692				& 0.5782				& 0.5718				& 0.5655 				\\
Profile + Network										 	 & 0.6923					& 0.6308				& 0.6205				& 0.6077				& 0.5986 	 \\
Text Similarity        								 & 0.5231					& 0.5154				& 0.5436			 	& 0.5231				& 0.5538 			\\
Profile       												 & 0.5846					& 0.5769				& 0.5897			 	& 0.5923 				& 0.5895  		\\
Network        										 		 & 0.6462					& 0.6462  			& 0.6121				& 0.6128				& 0.5990   \\
\hline
Expert Finding (Yang et al.)~\cite{yang09bole} &	0.5500	& 0.6000			& ~~0.6333			&~~--						&~~0.6356   \\
\hline
\end{tabular}
}
\caption{The results obtained with the different sets of features.}
\label{t3}
\end{table}

Since DBLP has rich information about citation links, we can see that the set of network features achieve the best results for this dataset in terms of MAP. The results also show that, individually, textual similarity features have the poorest results. This means that considering only textual evidence provided by query topics, together with article's titles and abstracts, may not be enough to determine if some authors are experts or not, and that indeed the information provided by citation and co-authorship patterns can help in expert retrieval. Finally, when comparing our unsupervised method against the supervised learning to rank approach proposed by Yang et al.~\cite{yang09bole}, showing that our approach provides very competitive results against the supervised method. Notice that unsupervised approaches are particularly interesting in the context of expert search systems for academic digital libraries, since relevance judgments for specific areas of knowledge, which are required to the usage of supervised approaches, are hard to obtain.

\section{Conclusions}

This paper argued that unsupervised rank aggregation methods provide a sound approach for combining multiple estimators of expertise, derived from the textual contents, from the graph-structure of the community of experts, and from expert profile information. Experiments on a dataset of academic publications show very competitive results in terms of P@5 and MAP, attesting for the adequacy of the proposed approaches. This is particularly interesting to the application domain of academic expert search, since the relevance judgments required by supervised approaches are only scarcely available.

Despite the interesting results, there are also many ideas for future work. Recent works have, for instance, proposed that there are advanced unsupervised rank aggregation methods capable of outperforming CombSUM and CombMNZ. This is currently a very hot topic of research and, for
future work, we would for instance like to experiment with the ULARA algorithm recently proposed by Klementiev et al.~\cite{Klementiev09Unsupervised} or explore supervised approached based on Learning to Rank techniques as the ones that have been explored in the field o Question/Answering~\cite{Moreira11iva}.

\bibliographystyle{plain}

\begin{thebibliography}{10}

\bibitem{Pine11Data}
Julien Ah-Pine.
\newblock On data fusion in information retrieval using different aggregation
  operators.
\newblock {\em Web Intelligence and Agent Systems}, 9(1), 2011.

\bibitem{Balog06Formal}
K.~Balog, L.~Azzopardi, and M.~de~Rijke.
\newblock Formal models for expert finding in enterprise corpora.
\newblock In {\em Proceedings of the 29th annual international {ACM SIGIR}
  conference on Research and development in information retrieval}, 2006.

\bibitem{Banks06extension}
M.~Banks.
\newblock An extension of the {H}irsch index: Indexing scientific topics and
  compounds.
\newblock {\em Scientometrics}, 69(1), 2006.

\bibitem{Batista06Possible}
P.~D. Batista, M.~G. Campiteli, O.~Kinouchi, and A.~S. Martinez.
\newblock Is it possible to compare researchers with different scientific
  interests?
\newblock {\em Scientometrics}, 68(1), 2006.

\bibitem{Brin99PageRank}
S.~Brin, L.~Page, R.~Motwani, and T.~Winograd.
\newblock The pagerank citation ranking: Bringing order to the web.
\newblock Technical Report 1999-66, Stanford Digital Library Technologies
  Project, 1999.

\bibitem{Cao06Research}
Y.~Cao, J.~Liu, S.~Bao, and H.~Li.
\newblock Research on expert search at enterprise track of {TREC} 2005.
\newblock In {\em Proceedings of the 14th Text REtrieval Conference}, 2006.

\bibitem{Chen07Finding}
P.~Chen, H.~Xie, S.~Maslov, and S.~Redner.
\newblock Finding scientific gems with {G}oogle's page rank algorithm.
\newblock {\em Journal of Informetrics}, 1(1), 2007.

\bibitem{Craswell06Overview}
N.~Craswell, A.~P. de~Vries, and I.~Soboroff.
\newblock Overview of the {TREC}-2005 enterprise track.
\newblock In {\em Proceedings of the 14th Text REtrieval Conference}, 2006.

\bibitem{Deng08Formal}
H.~Deng, I.~King, and M.~R. Lyu.
\newblock Formal models for expert finding on {DBLP} bibliography data.
\newblock In {\em Proceedings of the 8th {IEEE} International Conference on
  Data Mining}, 2008.

\bibitem{Dwork01Rank}
Cynthia Dwork, Ravi Kumar, Moni Naor, and D.~Sivakumar.
\newblock Rank aggregation methods for the web.
\newblock In {\em Proceedings of the 10th international conference on World
  Wide Web}, 2001.

\bibitem{Egghe06Theory}
L.~Egghe.
\newblock Theory and practise of the $g$-index.
\newblock {\em Scientometrics}, 69(1), 2006.

\bibitem{Fang07Probabilistic}
H.~Fang and C.~Zhai.
\newblock Probabilistic models for expert finding.
\newblock In {\em Proceedings of the 29th European Conference on Information
  Retrieval Research}, 2007.

\bibitem{Farah07Outranking}
Mohamed Farah and Daniel Vanderpooten.
\newblock An outranking approach for rank aggregation in information retrieval.
\newblock In {\em Proceedings of the 30th annual international ACM SIGIR
  conference on Research and development in information retrieval}, 2007.

\bibitem{Fox94Combination}
E.~A. Fox and J.~A. Shaw.
\newblock Combination of multiple searches.
\newblock In {\em Proceedings of the 2nd Text Retrieval Conference}, 1994.

\bibitem{Hirsch05Index}
J.~E. Hirsch.
\newblock An index to quantify an individual's scientific research output.
\newblock {\em Proceedings of the National Academy of Sciences USA}, 102(46),
  2005.

\bibitem{Klementiev09Unsupervised}
Alexandre Klementiev, Dan Roth, Kevin Small, and Ivan Titov.
\newblock Unsupervised rank aggregation with domain-specific expertise.
\newblock In {\em Proceedings of the 21st International Joint Conference on
  Artifical intelligence}, 2009.

\bibitem{Liu05authorship}
X.~Liu, J.~Bollen, M.~L. Nelson, and H.~Van de~Sompel.
\newblock Co-authorship networks in the digital library research community.
\newblock {\em Information Processing and Management}, 41(6), 2005.

\bibitem{Macdonald08Voting}
C.~Macdonald and I.~Ounis.
\newblock Voting techniques for expert search.
\newblock {\em Knowledge and Information Systems}, 16(3), 2008.


\bibitem{Moreira11MSc}
C.~Moreira.
\newblock Learning to Rank Academic Experts.
\newblock {\em  Master Thesis}, Technical University of Lisbon, 2011.

\bibitem{Moreira11iva}
C.~Moreira, A.~C.~Mendes, L.~Coheur, and B~Martins.
\newblock Towards the rapid development of a natural language understanding module.
\newblock In {\em Proceedings of the 10th International on Intelligent Virtual Agents}, 2011.

\bibitem{Petkova07Proximity}
D.~Petkova and W.~B. Croft.
\newblock Proximity-based document representation for named entity retrieval.
\newblock In {\em Proceedings of the 16th {ACM} Conference on Information and
  Knowledge Management}, 2007.

\bibitem{Pavel09Search}
P.~Serdyukov.
\newblock {\em Search for Expertise : Going Beyond Direct Evidence}.
\newblock PhD thesis, University of Twente, 2009.

\bibitem{Antonis06Generalized}
A.~Sidiropoulos, D.~Katsaros, and Y.~Manolopoulos.
\newblock Generalized $h$-index for disclosing latent facts in citation
  networks.
\newblock {\em Scientometrics}, 2006.

\bibitem{Sidiropoulos05citation}
A.~Sidiropoulos and Y.~Manolopoulos.
\newblock A citation-based system to assist prize awarding.
\newblock {\em ACM SIGMOD Record}, 34(4), 2005.

\bibitem{Sidiropoulos06generalized}
A.~Sidiropoulos and Y.~Manolopoulos.
\newblock Generalized comparison of graph-based ranking algorithms for
  publications and authors.
\newblock {\em Journal for Systems and Software}, 79(12), 2006.

\bibitem{Soboroff07Overview}
I.~Soboroff, A.~P. de~Vries, and N.~Craswell.
\newblock Overview of the {TREC}-2006 enterprise track.
\newblock In {\em Proceedings of the 15th Text REtrieval Conference}, 2007.

\bibitem{Zhu07Open}
J.~Zhu~S. Song, S.~R{\"u}ger, M.~Eisenstadt, and E.~Motta.
\newblock The open university at {TREC 2006} enterprise track expert search
  task.
\newblock In {\em Proceedings of the 15th Text REtrieval Conference}, 2007.

\bibitem{Zhu08Modeling}
J.~Zhu~S. Song, S.~R{\"u}ger, and J.~Huang.
\newblock Modeling document features for expert finding.
\newblock In {\em Proceedings of the 17th {ACM} Conference on Information and
  Knowledge Management}, 2008.

\bibitem{yang09bole}
Y.~Yang, J.~Tang, B.~Wang, J.~Guo, J.~Li, and S.~Chen.
\newblock Expert2{B}ole: From expert finding to bole search.
\newblock In {\em Knowledge Discovery and Data Mining}, 2009.

\bibitem{ZhangEIndex}
Chun-Ting Zhang.
\newblock The e-index, complementing the h-index for excess citations.
\newblock {\em Public Library of Science One}, 4, 2009.

\end{thebibliography}

\end{document}